\documentstyle[twocolumn,aps,pra,epsfig,subeqn]{revtex}

\newcommand{\beq}{\begin{equation}}
\newcommand{\eeq}{\end{equation}}
\newcommand{\eq}[1]{Eq. (\ref{#1})}

\begin{document}
\title{Dynamics of a two-mode Bose-Einstein condensate beyond mean field theory}
\author{J.R. Anglin and A. Vardi}
\address{ITAMP, Harvard-Smithsonian Center for Astrophysics\\
60 Garden Street, Cambridge MA 02138}
\date{\today}
\maketitle

\begin{abstract}
We study the dynamics of a two-mode Bose-Einstein condensate in the vicinity
of a mean-field dynamical instability. Convergence to mean-field theory
(MFT), with increasing total number of particles $N$, is shown to be
logarithmically slow. Using a density matrix formalism rather than the
conventional wavefunction methods, we derive an improved set of equations of
motion for the mean-field plus the fluctuations, which goes beyond MFT and
provides accurate predictions for the leading quantum corrections and the
quantum break time.  We show that the leading quantum corrections appear as
decoherence of the reduced single-particle quantum state; we also compare
this phenomenon to the effects of thermal noise. Using the rapid dephasing
near an instability, we propose a method for the direct measurement of
scattering lengths.
\end{abstract}

\section{Introduction}

The effective low-energy Hamiltonian for $N$ interacting bosons confined in
an external potential $V_{ext}$, is given in second-quantized form as 
\begin{equation}
\hat{H}=\int \!d^{3}r{\bf \,}{\hat{\psi}}^{\dagger }\left[ -\frac{\hbar ^{2}%
}{2m}\nabla ^{2}+V({\bf r})+\frac{g}{2}{\hat{\psi}}^{\dagger }{\hat{\psi}}%
\right] {\hat{\psi}}
\end{equation}
where $V({\bf r})$ is the external trapping potential, and $m$ is the
particle mass, $g$ is a coupling constant proportional to the s-wave
scattering length, and ${\hat{\psi}},\hat{\psi}^\dagger$ are bosonic 
annihilation and destruction operator fields obeying the canonical commutation 
relation 
$\left[ {\hat{\psi}}({\bf r}),{\hat{\psi}}^{\dagger }({\bf r}^{\prime })\right] 
=\delta ({\bf r}-{\bf r}^{\prime })$. 
(This Hamiltonian
is an effective low-energy approximation, in the sense that short wavelength
degrees of freedom have been eliminated: it is applicable in the regime of
ultracold scattering, where short distance modes are only populated
virtually, during brief two-body collisions.) \ At very low temperatures,
Bose-Einstein condensation occurs, so that a large fraction of the particles
occupy the same single-particle state, characterized by the single particle
wave function $\Psi ({\bf r},t).$ \ In this regime one can formulate a
perturbative expansion in the small quantity $N^{-1/2}$, where $N$ is the
number of particles in the condensate, whose result at leading order is the
Gross-Pitaevskii nonlinear Schr\"{o}dinger equation (GPE) governing the
condensate wave function:
\begin{equation}
i\hbar \frac{\partial }{\partial t}\Psi ({\bf r},t)=\left( -\frac{\hbar
^{2}\nabla ^{2}}{2m}+V_{ext}({\bf r})+g|\Psi ({\bf r},t)|^{2}\right) \Psi (%
{\bf r},t)\;,
\end{equation}

The Gross-Pitaevskii mean field theory (MFT) provides a classical field
equation for nonlinear matter waves, which is generally considered as `the
classical limit' of the Heisenberg equation of motion for the field operator 
${\hat{\psi}}$ (which is of precisely the same form). \ We can make precise
the sense in which it is a classical limit, by reformulating the system
governed by (\ref{ham}) in the path integral representation. \ We will not
actually use this formulation in this paper; we merely note the GPE is the
saddlepoint equation that appears in a steepest descents approximation to
the path integral. \ This is precisely the standard semi-classical
approximation, with the exception that $1/N$ is playing the role usually
played by $\hbar $. \ Hence despite the resemblance of the GPE to a Schr\"{o}dinger equation, 
complete with finite $\hbar$, we can indeed identify MFT
as the classical limit, in essentially the same sense as in the case $\hbar
\rightarrow 0$, of the quantum field theory. \ Because $N$ in current
trapped dilute alkali gas BEC experiments is characteristically large
(typically of the order $10^{5}-10^{8}$ atoms), qualitatively significant
quantum corrections to MFT are hard to observe, and the GP theory is highly
successful in predicting experimental results. 

The entire field of quantum chaos is founded upon one property of the
classical limit, however, which is that convergence to classicality as 
$\hbar \rightarrow 0$\ is logarithmically slow if classical trajectories
diverge exponentially. \ This implies that we must expect
strong quantum corrections to MFT in the vicinity of a dynamically unstable
fixed point. \ In particular, the quantum evolution will depart
significantly from the classical approximation after a logarithmic `quantum
break time', which will be $\symbol{126}\log N$ in our case, as it is $%
\symbol{126}\log (1/\hbar )$ in the standard case . \ \ In our case, the
nature of this departure is that after the quantum break time, a condensate
will become significantly depleted, as exponential production of
quasi-particles transfers particles to orthogonal modes \cite{CastinDum}. \
Depletion of the condensate means, by definition, that the single particle
reduced density matrix (SPDM) becomes quantum mechanically less pure. \
Hence for a condensate, just as the classical limit of the quantum field
theory resembles the quantum mechanics of a single particle, so quantum
corrections at the field theory level appear as quantum decoherence in the
single-particle picture. \ Since decoherence is most often considered as
enforcing classicality, there is something like irony in this situation. \
And it suggests that studying the corrections to MFT for Bose-Einstein
condensates may give us some new insights into decoherence; and that some
aspects of decoherence may be useful in understanding condensates beyond
MFT. \ \ This is the motivation for the work we now report.\ 

In this paper we provide the details of a previously published study \cite
{vardi} of the correspondence between mean-field and exact quantum dynamics
of a two-mode BEC. The model system contains an isolated dynamical
instability for certain regions of parameter space. We show that quantum
corrections in the vicinity of this unstable state, do indeed become
significant on a short $\log (N)$ time scale, whereas quantum effects in
other regions of phase space remain small $1/\sqrt{N}$ corrections. We
present a simple theory that goes beyond MFT and provides accurate
predictions of the leading quantum corrections, by taking one further step
in the so called Bogoliubov-Born-Green-Kirkwood-Yvon (BBGKY) hierarchy.  In
accordance with our view of quantum corrections as decoherence, we use a
density-matrix Bloch picture to depict the dephasing process. The
density-matrix formalism has the additional advantage of allowing for
initial conditions that are not covered by the Hartree-Fock-Bogoliubov
Gaussian ansatz, and which better correspond to the physical state of the
system.
\begin{figure}
\epsfig{file=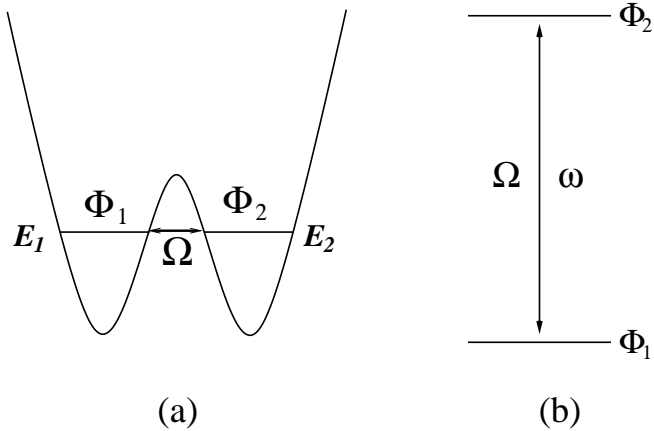,width=0.9\columnwidth,angle=-90}
\caption{Two-mode Bose-Einstein condensates: (a) a condensate in a
double-well potential (b) a spinor condensate.}
\label{f1}
\end{figure}
In section II we briefly review the model system and its experimental
realizations. In section III we derive the mean-field equations of motion in
the Bloch representation, and illustrate the main features of the produced
dynamics for various parameter sets. Quantum corrections to the two-mode MFT
are studied section IV, as well as an improved theory that predicts the
leading corrections. In section V we consider the effect of thermal noise,
and show an analogy between the quantum dephasing of the reduced single
particle density operator and thermal decoherence. In section VI we present
a potential application of the rapid decoherence near the dynamical
instability of the two mode model, for the measurement of s-wave
scattering-lengths. Discussion and conclusions are presented in section VII.

\section{The two-mode condensate}

We consider a BEC in which particles can only effectively populate either
one of two second-quantized modes. Two possible experimental realizations of
this model are illustrated in Fig. 1. The first (Fig 1a) is a condensate
confined in a double-well trap \cite{javanainen,walls,Janne,smerzi,leggett,lewenstein} 
which may be formed by
splitting a harmonic trap with a far off resonance intense laser sheet 
\cite{ketterle}. In this case single-particle tunneling provides a linear
coupling between the local mode solutions of the individual wells, 
which can in principle be tuned over a wide
range of strengths by adjusting the laser sheet intensity. The two-mode
regime is reached when the self-interaction energy $gn$ is small compared to
the spacing between the trap modes $\hbar\omega_{trap}$: 
\begin{equation}
gn=\frac{4\pi\hbar^2 a}{m}\frac{N}{4\pi l^3}\ll \frac{\hbar^2}{ml^2}%
=\hbar\omega_{trap}~,
\end{equation}
where $l$ is the characteristic trap size. Thus the two-mode condition is, 
\begin{equation}
l\gg N|a|~.  \label{tmcon}
\end{equation}
The two-mode condition (\ref{tmcon}) may be met by double-well traps with
characteristic frequencies of the order of 100 Hz, containing several
hundred particles. When constructed, larger traps will maintain the two-mode
limit at higher $N$.

The second experimental realization of a two-mode BEC is the effectively
two-component spinor condensate \cite{cornell1,cornell2} depicted in Fig.
1b. In this case the linear coupling between the modes is provided by a near
resonant radiation field \cite{cornell3,stenholm}. If collisions do not
change spin states, the nonlinear interactions between the particles depend on
three scattering lengths $a_{ij}$.  In realisations of spinor condensates it is easy
to ensure $a_{11}=a_{22}$ by symmetry, in which case the nonlinear interaction term 
becomes $\hat{H}_{int}=\int\!d^3r\,\hat{\cal H}_{int}({\bf r})$ for 
\begin{eqnarray}\label{hint}
\hat{\cal H}_{int}&\propto\sum_{i,j} a_{ij} \hat{\psi}_i^\dagger\hat{\psi}_i\hat{\psi}_j^\dagger\hat{\psi}_j\to&{a_{11}+a_{12}\over2}(\hat{\psi}_1^\dagger\hat{\psi}_1
+\hat{\psi}_2^\dagger\hat{\psi}_2)^2
\nonumber\\
 &&+{a_{11}-a_{12}\over2}(\hat{\psi}_1^\dagger\hat{\psi}_1-\hat{\psi}_2^\dagger\hat{\psi}_2)^2\;.
\end{eqnarray}
We can therefore define two healing lengths $\xi_\pm=1/\sqrt{\rho(a_{11}\pm a_{12})}$, where 
$\rho$ is the total density,
characterizing the effect on the spatial state of the condensate of the two nonlinear terms.  The
two-mode regime, in which the spatial state is fixed and essentially independent of the internal 
state, is reached when $\xi_-$ becomes larger than the sample size (its largest dimension).  
Since for available alkali gases all $a_{ij}$ differ only by a few percent, $\xi_-<<\xi_+$, and 
hence the two-mode regime 
can be reached with $N<10^{4}$ atoms in weak, nearly spherical traps ($\omega _{trap}\leq 100$
Hz).  Less isotropic traps obviously reach the two-mode regime only at smaller $N$. To extend the 
internal state two mode regime to larger $N$ we must make the trap weaker;
for fixed $N$, $\xi_-$ scales with sample size $L$ as $L^{3/2}$.   
Hence to ensure $L/\xi_-<1$ for fixed $N$ requires a sufficiently large (weak) trap. For 
Rb and Na experiments, whose lifetimes are limited by three-body collisions, the slowing down of
the two-mode dynamics at reduced total density should be more than compensated for by the extended 
condensate lifespan.   

In both realizations, the many-body Hamiltonian reduces in the two-mode
limit (and in the spinor realization also in the rotating-wave
approximation) to the form, 
\[
\hat{H}(t)=\frac{E_{1}+E_{2}}{2}({\hat{a}}_{1}^{\dagger }{\hat{a}}_{1}+{%
\hat{a}}_{2}^{\dagger }{\hat{a}}_{2})-\frac{\hbar \Omega }{2}\left( {\hat{a}}%
_{1}^{\dagger }{\hat{a}}_{2}+{\hat{a}}_{2}^{\dagger }{\hat{a}}_{1}\right) 
\]
\begin{equation}
+\hbar g\left[ ({\hat{a}}_{1}^{\dagger })^{2}{\hat{a}}_{1}^{2}+({\hat{a}}%
_{2}^{\dagger })^{2}{\hat{a}}_{2}^{2}\right] 
\end{equation}
where $E_{1}$ and $E_{2}$ are the two condensate mode energies, $\Omega $ is
the coupling strength between the modes, $g$ is the two-body interaction
strength, and ${\hat{a}}_{1},{\hat{a}}_{1}^{\dagger },{\hat{a}}_{2},{\hat{a}}%
_{2}^{\dagger }$ are particle annihilation and creation operators for the
two modes. The total number operator $\hat{N}\equiv {\hat{a}}_{1}^{\dagger }{%
\hat{a}}_{1}+{\hat{a}}_{2}^{\dagger }{\hat{a}}_{2}$ commuted with $\hat{H}$
and may be replaced with the c-number $N$. Writing the self-interaction
operators as $({\hat{a}}_{1}^{\dagger })^{2}{\hat{a}}_{1}^{2}+({\hat{a}}%
_{2}^{\dagger })^{2}{\hat{a}}_{2}^{2}=[\hat{N}^{2}+({\hat{a}}_{1}^{\dagger }{%
\hat{a}}_{1}-{\hat{a}}_{2}^{\dagger }{\hat{a}}_{2})^{2}]/2$ and discarding
c-number terms, we obtain the two-mode Hamiltonian 
\begin{equation}
\hat{H}=-\frac{\hbar \Omega }{2}\left( {\hat{a}}_{1}^{\dagger }{\hat{a}}_{2}+%
{\hat{a}}_{2}^{\dagger }{\hat{a}}_{1}\right) +\frac{\hbar g}{2}\left( {%
\hat{a}}_{1}^{\dagger }{\hat{a}}_{1}-{\hat{a}}_{2}^{\dagger }{\hat{a}}%
_{2}\right) ^{2}~.  \label{ham}
\end{equation}
We will take $g$ and $\omega $ to be positive, since the relative phase
between the two modes may be re-defined arbitrarily, and since without
dissipation the overall sign of $\hat{H}$ is insignificant.

\section{Two-mode mean-field theory in the Bloch representation}

The conventional wavefunction formalisms consider the evolution of ${\hat{a}}%
_{j}$ and its expectation value in a symmetry-breaking ansatz (where the
symmetry being broken is that associated with conservation of $N$). Instead,
we will examine the evolution of the directly observable quantities $\hat{a}
_{i}^{\dagger }\hat{a}_{j}$, whose expectation values define the reduced
single particle density matrix (SPDM) $R_{ij}\equiv \langle \hat{a}%
_{i}^{\dagger }\hat{a}_{j}\rangle /N$. Writing the Hamiltonian of Eq. (\ref
{ham}) in terms of the SU(2) generators, 
\begin{eqnarray}\label{lxyz}
{\hat{L}}_{x} &\equiv &\frac{{\hat{a}}_{1}^{\dagger }{\hat{a}}_{2}+{\hat{a}}%
_{2}^{\dagger }{\hat{a}}_{1}}{2}~,  \nonumber \\
{\hat{L}}_{y} &\equiv &\frac{{\hat{a}}_{1}^{\dagger }{\hat{a}}_{2}-{\hat{a}}%
_{2}^{\dagger }{\hat{a}}_{1}}{2i}~, \\
{\hat{L}}_{z} &=&\frac{{\hat{a}}_{1}^{\dagger }{\hat{a}}_{1}-{\hat{a}}%
_{2}^{\dagger }{\hat{a}}_{2}}{2}~,  \nonumber
\end{eqnarray}
we obtain 
\begin{equation}
\hat{H}=-\hbar \Omega {\hat{L}}_{x}+\frac{\hbar g}{2}{\hat{L}}_{z}^{2}\;.
\label{hamiltonian}
\end{equation}
The Heisenberg equations of motion for the three angular momentum operators
of Eq. (\ref{lxyz}) read 
\begin{eqnarray}\label{Ldot}
\frac{d}{dt}{\hat{L}}_{x}=-\frac{i}{\hbar }[{\hat{L}}_{x},H] &=&-\frac{g
}{2}({\hat{L}}_{y}{\hat{L}}_{z}+{\hat{L}}_{z}{\hat{L}}_{y})~,  \nonumber
\\
\frac{d}{dt}{\hat{L}}_{y}=-\frac{i}{\hbar }[{\hat{L}}_{y},H] &=&+\Omega {%
\hat{L}}_{z}+\frac{g}{2}({\hat{L}}_{x}{\hat{L}}_{z}+{\hat{L}}_{z}{\hat{L}%
}_{x})~, \\
\frac{d}{dt}{\hat{L}}_{z}=-\frac{i}{\hbar }[{\hat{L}}_{z},H] &=&-\Omega {%
\hat{L}}_{y}~.  \nonumber
\end{eqnarray}

Thus the expectation values of the first-order operators ${\hat{L}}_{i}$
depend not only on themselves, but also on the second-order moments $\langle 
{\hat{L}}_{i}{\hat{L}}_{j}\rangle $. Similarly, the time evolution of the
second-order moments depends on third-order moments; and so on.
Consequently, we obtain the BBGKY hierarchy of equations of motion for the
expectation-values, 
\begin{eqnarray}\label{BBGKY}
\frac{d}{dt}\langle {\hat{L}}_{i}\rangle  &=&f\left( \langle {\hat{L}}%
_{i^{\prime }}\rangle ,\langle {\hat{L}}_{i^{\prime }}{\hat{L}}_{j^{\prime
}}\rangle \right) ~,  \nonumber \\
\frac{d}{dt}\langle {\hat{L}}_{i}{\hat{L}}_{j}\rangle  &=&f\left( \langle {%
\hat{L}}_{i^{\prime }}{\hat{L}}_{j^{\prime }}\rangle ,\langle {\hat{L}}%
_{i^{\prime }}{\hat{L}}_{j^{\prime }}{\hat{L}}_{k^{\prime }}\rangle \right)
~, \\
\frac{d}{dt}\langle {\hat{L}}_{i}{\hat{L}}_{j}{\hat{L}}_{k}\rangle 
&=&f\left( \langle {\hat{L}}_{i^{\prime }}{\hat{L}}_{j^{\prime }}{\hat{L}}%
_{k^{\prime }}\rangle ,\langle {\hat{L}}_{i^{\prime }}{\hat{L}}_{j^{\prime }}%
{\hat{L}}_{k^{\prime }}{\hat{L}}_{l^{\prime }}\rangle \right) ~,  \nonumber
\\
~ &\vdots &~  \nonumber
\end{eqnarray}
where $i,j,k,\dots ,i^{\prime },j^{\prime },k^{\prime },l^{\prime },\dots
=x,y,z$. In order to obtain a closed set of equations of motion, the
hierarchy of Eq. (\ref{BBGKY}) must be truncated at some stage by
approximating the $N$-th order expectation value in terms of all lower-order
moments.

The lowest-order truncation of Eq. (\ref{BBGKY}) is obtained by
approximating the second-order expectation values $\langle {\hat{L}}_{i}{%
\hat{L}}_{j}\rangle $ as products of the first-order moments $\langle {\hat{L%
}}_{i}\rangle $ and $\langle {\hat{L}}_{j}\rangle $: 
\begin{equation}
\langle {\hat{L}}_{i}{\hat{L}}_{j}\rangle \approx \langle {\hat{L}}%
_{i}\rangle \langle {\hat{L}}_{j}\rangle ~.  \label{mfa}
\end{equation}
The equations of motion for the single-particle Bloch vector 
\begin{equation}
{\bf {\vec{s}}}=(s_{x},s_{y},s_{z})\equiv \left( \frac{2\langle {\hat{L}}%
_{x}\rangle }{N},\frac{2\langle {\hat{L}}_{y}\rangle }{N},\frac{2\langle {%
\hat{L}}_{z}\rangle }{N}\right) ,
\end{equation}
Then read 
\begin{eqnarray}\label{nlbloch}
{\dot{s}}_{x} &=&-\kappa s_{z}s_{y}~,  \nonumber \\
{\dot{s}}_{y} &=&\Omega s_{z}+\kappa s_{z}s_{x}~, \\
{\dot{s}}_{z} &=&-\Omega s_{y}~,  \nonumber
\end{eqnarray}
where $\kappa =gN/2$. Equations (\ref{nlbloch}) describe rotations of
the Bloch vector ${\bf {\vec{s}}}$, and so the norm $|{\bf {\vec{s}}}|$ is
conserved in MFT. Consequently, for a pure SPDM, Eq. (\ref{nlbloch}) are
completely equivalent to the two-mode Gross-Pitaevskii equation \cite{smerzi}%
, 
\begin{mathletters}
\begin{eqnarray}
i\frac{\partial }{\partial t}a_{1} &=&\kappa a_{1}-\Omega a_{2} \\
i\frac{\partial }{\partial t}a_{2} &=&\kappa a_{2}-\Omega a_{1}
\end{eqnarray}
where $a_{1}$ and $a_{2}$ are the c-number
coefficients replacing the creation and annihilation operators of Eq. (\ref
{lxyz})
\begin{figure}
\begin{center}
\epsfig{file=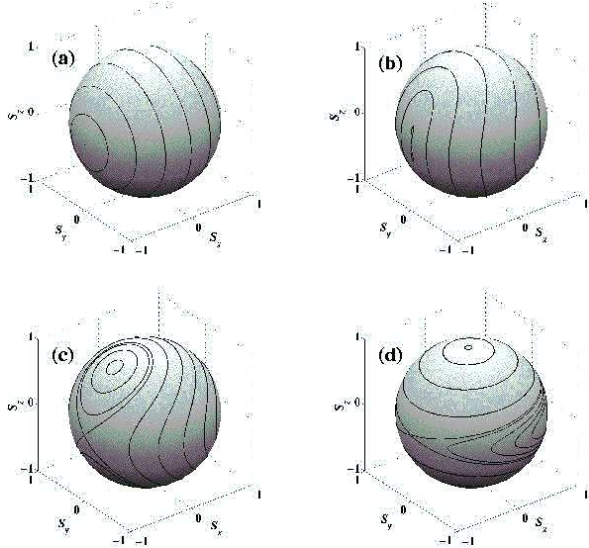,width=0.9\columnwidth}
\end{center}
\caption{ Mean-field trajectories at (a)$\protect\kappa=0$, (b)$\kappa=1.02\Omega$, 
(c)$\kappa=2\Omega$, and (d)$\kappa=20\Omega$. }
\label{f2}
\end{figure}
In Fig. 2 we plot mean-field trajectories at four different $\kappa/\Omega$
ratios. The nonlinear Bloch equations (\ref{nlbloch}) depict a competition
between linear Rabi oscillations in the $s_ys_z$-plane and nonlinear
oscillations in the $s_xs_y$-plane. For a noninteracting condensate (Fig.
2a) the trajectories on the Bloch sphere depict harmonic Rabi oscillations
about the $s_x$ axis. As $\kappa$ increases the oscillations become
increasingly anharmonic. As long as $\kappa<\Omega$ the nonlinearities may
be treated as perturbation. However, above the critical value $\kappa=\omega$
(Fig. 2b), there are certain regions in phase-space which are dominated by
the nonlinear term. The stationary point ${\bf {\vec s}}=(-1,0,0)$,
corresponding to the Josephson $\pi$-state (equal populations and a $\pi$
phase-difference), becomes dynamically unstable and the two trajectories
passing asymptotically close to it form a ``figure-eight''. The region
outside these limiting trajectories is dominated by the linear oscillations
whereas inside the nonlinear term prevails. Starting at the critical value
of $\kappa=2\Omega$ (Fig. 2c) population prepared in one of the modes
remains trapped in the half-sphere it originated from, conducting
oscillations with a non-vanishing time averaged population imbalance $%
\langle s_z\rangle_t\neq 0$. This phenomenon was termed ``macroscopic
self-trapping'' \cite{smerzi}. Finally, when $\kappa\gg\Omega$ (Fig. 2d) the
nonlinearity dominates the entire Bloch sphere, except for a narrow band
about the $s_z=0$ plain. 

\section{Quantum corrections and Bogoliubov backreaction}

In the vicinity of the dynamically unstable point, we expect MFT to break
down on a time scale only logarithmic in $N$. In order to verify this
prediction, we solve the full $N$-body problem exactly, by fixing the total
number of particles $N$, thereby restricting the available phase-space to
Fock states of the type $|\,{n,N-n}\,\rangle $ with $n$ particles in one
mode and $N-n$ particles in the other mode, $n$ ranging from 0 to $N$. Thus
we obtain an $N+1$ dimensional representation for the Hamiltonian (\ref
{hamiltonian}) and the $N$-body density operator $\hat{\rho}$: 
\end{mathletters}
\begin{eqnarray}
H_{m,n} &=&\langle \,{m,N-m}\,|\,\hat{H}\,|\,{n,N-n}\,\rangle ~,
\label{hnrep} \\
\rho _{m,n} &=&\langle \,{m,N-m}\,|\,\hat{\rho}\,|\,{n,N-n}\,\rangle ~,
\end{eqnarray}
for $m,n=0,1,\dots ,N$. The exact quantum solution is obtained numerically
by propagating $\hat{\rho}$ according to the Liouville von-Neumann equation 
\begin{equation}
i\hbar \dot{\hat\rho}=[\hat{H},\hat{\rho}]~.  \label{lvn}
\end{equation}
Using the Hamiltonian of Eq. (\ref{hamiltonian}) to evaluate the matrix
elements of Eq. (\ref{hnrep}) and substituting into Eq. (\ref{lvn}), we
obtain dynamical equations for the N-body density matrix: 
\begin{eqnarray}\label{rhonmdot}
i\hbar \dot{\rho}_{m,n} &=&-\frac{\Omega }{2}\left[ \sqrt{m(N-m+1)}\rho
_{m-1,n}\right.   \nonumber  
\\
~ &~&+\sqrt{(m+1)(N-m)}\rho _{m+1,n}\nonumber\\
~ &~&-\sqrt{n(N-n+1)}\rho _{m,n-1}\\
~ &~&\left.-\sqrt{(n+1)(N-n)}\rho _{m,n+1}\right]   \nonumber \\
~ &~&+\frac{g}{4}\left[ m^{2}-(N-m)^{2}-n^{2}+(N-n)^{2}\right] \rho
_{m,n}~.\nonumber
\end{eqnarray}
Equations (\ref{rhonmdot}) are solved numerically, using a Runge-Kutta
algorithm. In fig. 3 we plot exact quantum trajectories starting with all
particles in one mode, for increasingly large $N$ ($\kappa $ being fixed)
versus the corresponding mean-field trajectory. While MFT assumes a
persistently pure single particle state, quantum corrections to MFT appear
in the single-particle picture, as decoherence of the SPDM. When the
mean-field trajectory stays away from the instability (Fig. 3a) the quantum
trajectories indeed enter the interior of the unit Bloch sphere at a rate
that vanishes as $1/\sqrt{N}$. However, when the mean-field trajectory
includes the unstable state (Fig. 3b), we observe a sharp break of the
quantum dynamics from the mean-field trajectory at a time that only grows
slowly with $N$. 

In accordance with our picture of quantum corrections as decoherence, and in
order to obtain a more quantitative view of the entanglement-induced
dephasing process, we plot the von Neumann entropy 
\begin{equation}
S=Tr(R\ln R)=-\frac{1}{2}\ln \left[ \frac{(1+|{\bf {\vec{s}}}|)^{(1+|{\bf {%
\vec{s}}}|)}(1-|{\bf {\vec{s}}}|)^{(1-|{\bf {\vec{s}}}|)}}{4}\right] 
\end{equation}
of the exact reduced single-particle density operator, as a function of the
rescaled time $\Omega t$ for the same initial conditions as in Fig. 3. The
results are shown in Fig. 4. Since the entropy of mean-field trajectories is
identically zero, $S$ may serve as a measure of the deviation from MFT. When
the mean-field trajectory is stable (Fig. 4a), the single particle entropy
grows at a steady rate which vanishes as $N$ is increased. The variations in
the entropy growth curve are a function of the distance from the
instability. Near the instability (Fig. 4b) quantum corrections grow rapidly
at a rate which is independent of $N,$ and the time at which this divergence
takes place (the quantum break time) evidently grows only as $\log (N)$.
\begin{figure}
\centerline{
\epsfig{file=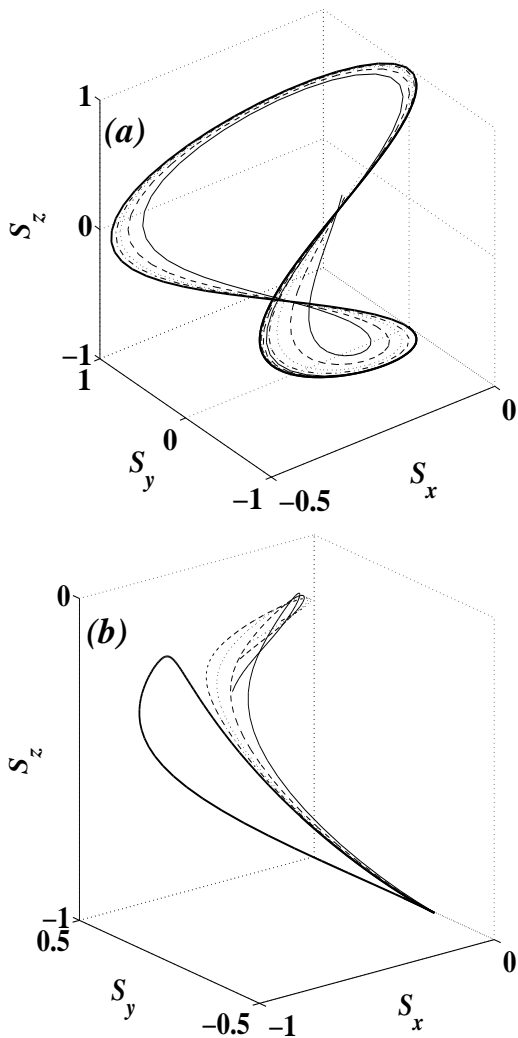,width=0.8\columnwidth}}
\label{f3}
\caption{Exact quantum trajectories starting with all particles in one mode,
with $N=50$ (-----), 100 ($---$), 200 ($\cdots $) and 400 ($-\cdot -$)
particles vs. the corresponding mean-field trajectory ({\bf -----}) for (a) 
$\kappa =\Omega $ and (b) $\kappa =2\Omega$.}
\end{figure}
Since MFT can thus easily fail near dynamical instabilities, it is highly
desirable to obtain an improved theory in which Bloch-space trajectories
would be allowed to penetrate into the unit sphere without having to
simulate the entire $N$-body dynamics. In fact, such an improved non-unitary
theory is easily derived using the next level of the BBGKY hierarchy (\ref
{BBGKY}). This hierarchy truncation approach is in fact a systematic perturbative
approximation; but it is state-dependent.  That is, it provides a perturbative
approximation, not to the general evolution, but to the evolution of a special class
of initial states, within which the perturbative parameter is small.  In the case
of ultracold bosons, the phenomenon of Bose-Einstein condensation ensures that there
is a commonly realisable class of states in which the system is a 
{\it mildly fragmented condensate}. In our two mode model, this means that
the two eigenvalues of $R$ are $f$ and 1-$f$ for $f\ll 1$; and so from such initial
states we can approximate the evolution perturbatively using $f$ as our small 
parameter.  

\begin{figure}
\centerline{\epsfig{file=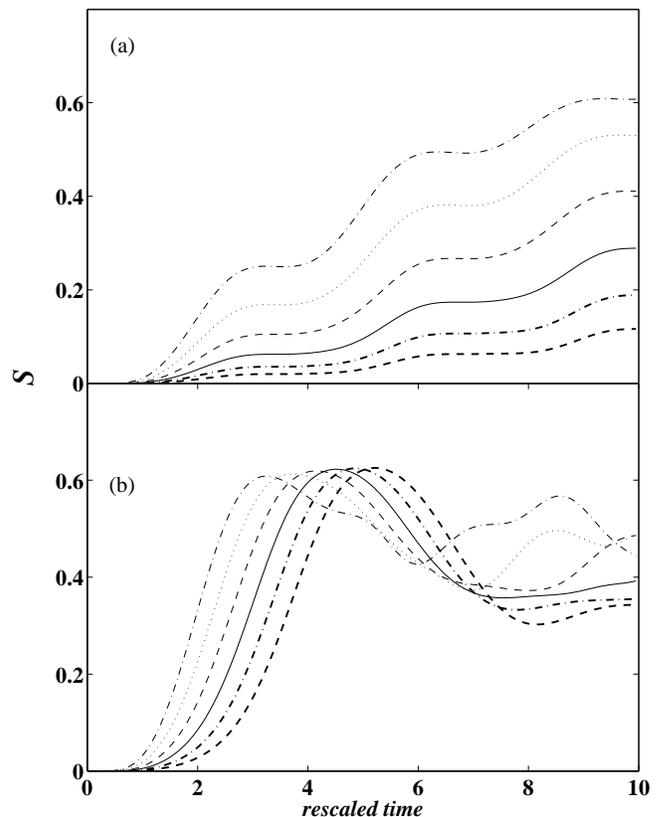,width=\columnwidth}}
\caption{Growth of the von Neumann entropy $S$ of the quantum reduced
single-particle density operator, at (a) $\protect\kappa =\Omega $ and (b) 
$\kappa =2\Omega $, for N=10 ($-\cdot -$), 20 ($\cdot \cdot \cdot $),
40 ($---$), 80 (------), 160 ({\bf - . - . -}), and 320 ({\bf - - - -})
particles. Initial conditions are the same as in Fig. 3. }
\label{f4}
\end{figure}
To zeroth order in $f$, $R$ is by definition a pure state, and hence
we have the MFT evolution on the surface of the Bloch sphere.  Going to next order in $f$
can be achieved by truncating the BBGKY hierarchy at one order higher.
We take ${
\hat{L}}_{i}=L_{i}+\hat{\delta L_{i}}$, where the c-number $L_{i}$ is ${\cal %
O}(N)$ and all the matrix elements of $\hat{\delta L_{i}}$ remain smaller
than ${\cal O}(N\sqrt{f})$ throughout the evolution of the system. The
second order moments, 
\begin{equation}
\Delta _{ij}=4N^{-2}\left( \langle {\hat{L}}_{i}{\hat{L}}_{j}+{\hat{L}}_{j}{%
\hat{L}}_{i}\rangle -2\langle {\hat{L}}_{i}\rangle \langle {\hat{L}}%
_{j}\rangle \right) ~,  \label{dij}
\end{equation}
will then be of order $f$. Writing the Heisenberg equations of motion for
the first- and second-order operators ${\hat{L}}_{i},{\hat{L}}_{i}{\hat{L}}%
_{j}$, taking their expectation values and truncating Eq. (\ref{BBGKY}) by
approximating 
\[
\langle {\hat{L}}_{i}{\hat{L}}_{j}{\hat{L}}_{k}\rangle \approx \langle {\hat{%
L}}_{i}{\hat{L}}_{j}\rangle \langle {\hat{L}}_{k}\rangle +\langle {\hat{L}}%
_{i}\rangle \langle {\hat{L}}_{j}{\hat{L}}_{k}\rangle +\langle {\hat{L}}_{i}{%
\hat{L}}_{k}\rangle \langle {\hat{L}}_{j}\rangle 
\]
\begin{equation}
-2\langle {\hat{L}}_{i}\rangle \langle {\hat{L}}_{j}\rangle \langle {\hat{L}}%
_{k}\rangle \;.  \label{trunc2}
\end{equation}
instead of the mean-field approximation (\ref{mfa}), we obtain the following
set of nine equations for the first- and second-order moments: 
\begin{eqnarray}\label{bgb} 
{\dot{s}}_{x} &=&-\kappa s_{z}s_{y}-\frac{\kappa }{2}\Delta _{yz}  \nonumber\\
{\dot{s}}_{y} &=&\omega s_{z}+\kappa s_{z}s_{x}+\frac{\kappa }{2}\Delta _{xz} 
\nonumber \\
{\dot{s}}_{z} &=&-\omega s_{y}  \nonumber \\
{\dot{\Delta}}_{xz} &=&-\omega \Delta _{xy}-\kappa s_{z}\Delta _{yz}-\kappa s%
_{y}\Delta _{zz}  \nonumber \\
{\dot{\Delta}}_{yz} &=&\omega (\Delta _{zz}-\Delta _{yy})+\kappa s_{z}\Delta
_{xz}+\kappa s_{x}\Delta _{zz} \\
{\dot{\Delta}}_{xy} &=&(\omega +\kappa s_{x})\Delta _{xz}-\kappa s_{y}\Delta
_{yz}+\kappa s_{z}(\Delta _{xx}-\Delta _{yy})  \nonumber \\
{\dot{\Delta}}_{xx} &=&-2\kappa s_{y}\Delta _{xz}-2\kappa s_{z}\Delta _{xy} 
\nonumber \\
{\dot{\Delta}}_{yy} &=&2(\omega +\kappa s_{x})\Delta _{yz}+2\kappa s_{z}\Delta
_{xy}  \nonumber \\
{\dot{\Delta}}_{zz} &=&-2\omega \Delta _{yz}~.  \nonumber
\end{eqnarray}

Equations (\ref{bgb}) will be referred to as the ``Bogoliubov backreaction
equations'' (BBR), because they demonstrate how the mean-field Bloch vector ${\bf {%
\vec{s}}}$ drives the fluctuations $\Delta _{ij}$ -- which is the physics described
by the Bogoliubov theory of linearized quantum corrections to MFT; but they also make
the Bloch vector subject in turn
to backreaction from the fluctuations, via the coupling terms $-\kappa \Delta _{yz}/2$
and $\kappa \Delta _{xz}/2$. This back-reaction has the effect of breaking
the unitarity of the mean-field dynamics. Consequently, the BBR trajectories
are no longer confined to the surface of the Bloch sphere, but penetrate to
the interior (representing mixed-state $R_{ij}$, with two non-zero
eigenvalues).  (Obviously, if the trajectories penetrate the sphere too deeply, so 
that the smaller eigenvalue $f$ ceases to be small, the entire approach of 
perturbing in $f$ will break down.)

In order to demonstrate how the BBR equations (\ref{bgb}) improve on MFT, we
compare trajectories obtained by these two formalisms to the exact
50-particle trajectories of Fig. 3. Both the $\kappa =\Omega $ stable
mean-field trajectory and the $\kappa =2\Omega $ unstable mean-field
trajectory cases are plotted in Fig. 5a and Fig. 5b, respectively. The
initial conditions for the BBR equations are determined by the initial state 
$|\,{N,0}\,\rangle $ to be 
\begin{eqnarray}\label{bbrinit}
s_{z} &=&-1,  \nonumber \\
\Delta _{xx} &=&\Delta _{yy}=2/N, \\
s_{x} &=&s_{y}=\Delta _{xy}=\Delta _{xz}=\Delta _{yz}=\Delta _{zz}=0~. 
\nonumber
\end{eqnarray}

\begin{figure}
\centerline{\epsfig{file=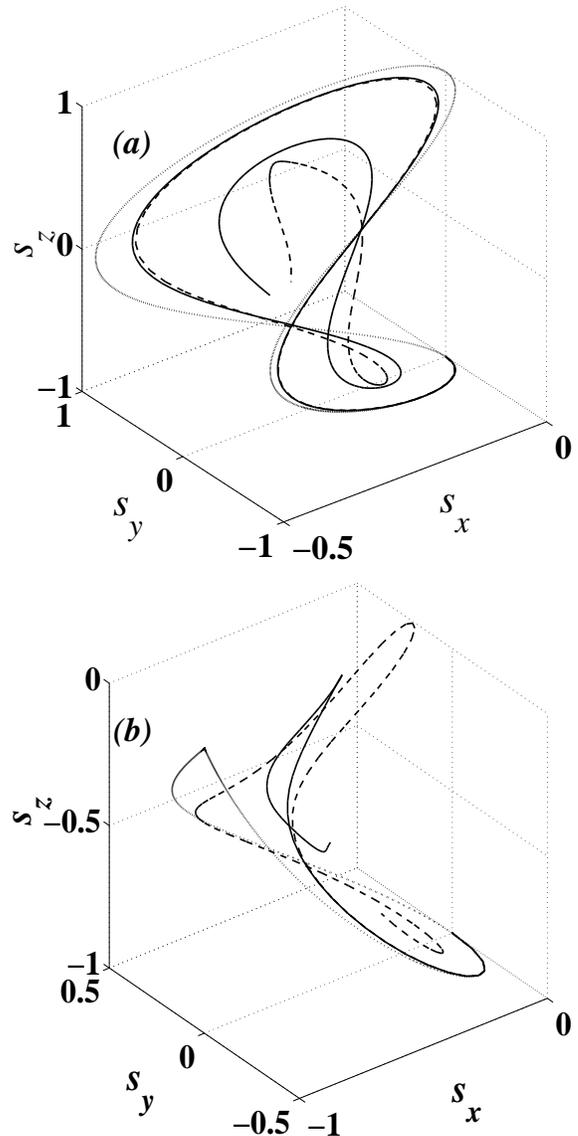,width=0.9\columnwidth}}
\caption{Mean field ($\cdot \cdot \cdot $), Bogoliubov back-reaction ($---$)
and exact 50 particles (-----) trajectories starting with all particles in
one mode, at (a) $\protect\kappa =\Omega $ and (b) $\protect\kappa =2\Omega$.}
\label{f5}
\end{figure}
The approximation of \eq{trunc2} ignores terms smaller than  
${\cal{O}}(f^{3/2})$. It is therefore better than the mean-field 
approximation (\ref{mfa}) by a factor of $f^{1/2}$. Consequently,  
as is clearly evident from Fig. 5a, the BBR equations (\ref{bgb}) are
far more successful than the mean-field equations (\ref{nlbloch}) in tracing
the full quantum dynamics. However, for any realistic number of particles,
the improvement is hardly necessary, as MFT would be accurate for very long
times. On the other hand, when the mean-field trajectory approaches the
instability (Fig. 5b), the BBR theory provides an accurate prediction of the
leading quantum corrections. Of course, since the BBR equations account for
only six moments unobserved by MFT, the period of the still quasiperiodic
BBR motion is shorter than that of the exact evolution and the BBR
trajectory eventually deviates from the quantum trajectory. Nevertheless,
the BBR formalism provides a simple and efficient method to predict the
quantum break time in large ($> 1000$ particles) condensates, for which full 
$N$-body simulations are restricted by available computation power.

The BBR equations (\ref{bgb}) are in fact identical to the equations of
motion one would obtain, for the same quantities, using the
Hartree-Fock-Bogoliubov Gaussian ansatz, in which second-order moments $%
\Delta_{ij}$ are initially factorized as $\Delta_{ij}=\delta_i\delta_j$ ($%
i,j=x,y,z$). Using this ansatz, the factorization persists and the time
evolution of $\delta_x,\delta_y$, and $\delta_z$ is equivalent to that of
perturbations of the mean-field equations (\ref{nlbloch}): 
\begin{eqnarray}  \label{bog2}
{\dot\delta}_x&=&-\kappa(s_z\delta_y+s_y\delta_z)~,  \nonumber \\
{\dot\delta}_y&=&\omega\delta_z+\kappa(s_z\delta_x+s_x\delta_z)~, \\
{\dot\delta}_z&=&-\omega\delta_y~.  \nonumber
\end{eqnarray}
Thus our equations for $\Delta_{ij}$ are in a sense equivalent to the usual
Bogoliubov equations. The quantitative advantage of our approach therefore
lies entirely in the wider range of initial conditions that it admits, which
may more accurately represent the exact initial conditions. For instance, a
Gaussian approximation will have $\Delta_{xx}={\cal O}(1)$ in the ground
state, where in fact $\Delta_{xx} = {\cal O}(N^{-1})$. This leads to an
error of order $N^{-1/2}$ in the Josephson frequency computed by linearizing
(\ref{bgb}) around the ground state, even though the Gaussian backreaction
result should naively be accurate at this order. Our SPDM approach does not
have this flaw, which is presumably the two-mode version of the
Hartree-Fock-Bogoliubov spectral gap \cite{griffin}.

\section{Dephasing due to thermal noise}

Decoherence is generally considered as suppressing quantum effects \cite
{giulini}. Ironically, in our case the leading quantum corrections to the
effectively classical MFT, are themselves decoherence of the single-particle
state of the condensate. Therefore, it is interesting to study the effect of
a realistic decoherence process, originating in the coupling to a bath of
unobserved degrees of freedom, on the interparticle entanglement process,
described in the previous section.

The main source of decoherence in BEC's is the thermal cloud of particles
surrounding the condensate. Thermal particles scattering off the condensate
mean field will for example, cause phase diffusion \cite{Anglin} at a rate $%
\Gamma$ proportional to the thermal cloud temperature. For internal states
not entangled with the condensate spatial state, $\Gamma$ may be as low as $%
10^{-5}$ Hz under the coldest experimental conditions, whereas for a double
well the rate may reach $10^{-1}$ Hz. Further sources of decoherence may be
described phenomenologically with a larger $\Gamma$. 

We account effect of thermal noise on the two-mode dynamics by using the
quantum kinetic master equation \cite{Janne}, 
\begin{equation}  \label{master}
\dot\rho=\frac{i}{\hbar}[\rho,H]-\frac{\Gamma}{2}\sum_{j=1,2} \left[{\hat a}%
_j^\dagger{\hat a}_j,\left[{\hat a}_j^\dagger{\hat a}_j,\rho\right]\right]~.
\end{equation}

Once again, we solve for ${\bf {\vec s}}(t)$ using either one of three
methods:

\noindent {\bf (a) MFT} - The decoherence term in Eq. (\ref{master})
introduces an exact $T_2=1/\Gamma$ transversal relaxation term into the
mean-field equations of motion: 
\begin{eqnarray}  \label{mfdec}
{\dot s}_x&=&-\kappa s_zs_y-\Gamma s_x ~,  \nonumber \\
{\dot s}_y&=&\omega s_z+\kappa s_zs_x-\Gamma s_y ~, \\
{\dot s}_z&=&-\omega s_y ~.  \nonumber
\end{eqnarray}

\noindent {\bf (b) BBR} - Evolving the first- and second-order operators
according to Eq. (\ref{master}), taking their expectation values and
truncating the hierarchy at the next level, we obtain the modified BBR
equations, 
\begin{eqnarray}\label{bbrdec}
{\dot s}_x&=&-\kappa s_zs_y-\frac{\kappa}{2}\Delta_{yz}-\Gamma s_x,  \nonumber
\\
{\dot s}_y&=&\omega s_z+\kappa s_zs_x+\frac{\kappa}{2}\Delta_{xz}-\Gamma s_y, 
\nonumber \\
{\dot s}_z&=&-\omega s_y ,  \nonumber \\
{\dot\Delta}_{xz}&=&-\omega\Delta_{xy}-\kappa s_z\Delta_{yz}-\kappa s%
_y\Delta_{zz}-\Gamma\Delta_{xz},  \nonumber \\
{\dot\Delta}_{yz}&=&\omega(\Delta_{zz}-\Delta_{yy})+\kappa s_z\Delta_{xz}+%
\kappa s_x\Delta_{zz}-\Gamma\Delta_{yz}, \\
{\dot\Delta}_{xy}&=&(\omega+\kappa s_x)\Delta_{xz}  \nonumber \\
~&~&-\kappa s_y\Delta_{yz}+\kappa s_z(\Delta_{xx}-\Delta_{yy})-4\Gamma(%
\Delta_{xy}+s_xs_y),  \nonumber \\
{\dot\Delta}_{xx}&=&-2\kappa s_y\Delta_{xz}-2\kappa s_z\Delta_{xy}-2\Gamma(%
\Delta_{xx}-\Delta_{yy}-2s_y^2),  \nonumber \\
{\dot\Delta}_{yy}&=&2(\omega+\kappa s_x)\Delta_{yz},  \nonumber \\
~&~&+2\kappa s_z\Delta_{xy}-2\Gamma(\Delta_{yy}-\Delta_{xx}-2s_x^2), 
\nonumber \\
{\dot\Delta}_{zz}&=&-2\omega\Delta_{yz}~,  \nonumber
\end{eqnarray}

\noindent {\bf (c) Exact quantum solution} - obtained by numerically
propagating the full $N$-particle density matrix under Eq. (\ref{master}).

In Fig. 6 we compare the Von-Neumann entropy of the exact $N$-body density
operator as a function of time for exponentially increasing $N$, to the
mean-field entropy. Due to the thermal noise, mean-field trajectories are no
longer confined to the zero-entropy sphere. However, whereas the quantum
break time in the absence of thermal noise has grown as $\log(N)$ (see Fig.
4b), it is clear from Fig. 6 that in the presence of this dephasing
mechanism it saturates to a finite value. Thus, while we may have naively
expected decoherence to reduce quantum corrections and thereby improve MFT,
in fact the addition of thermal dephasing has significantly damaged
classical-quantum correspondence.

\begin{figure}
\begin{center}
\epsfig{file=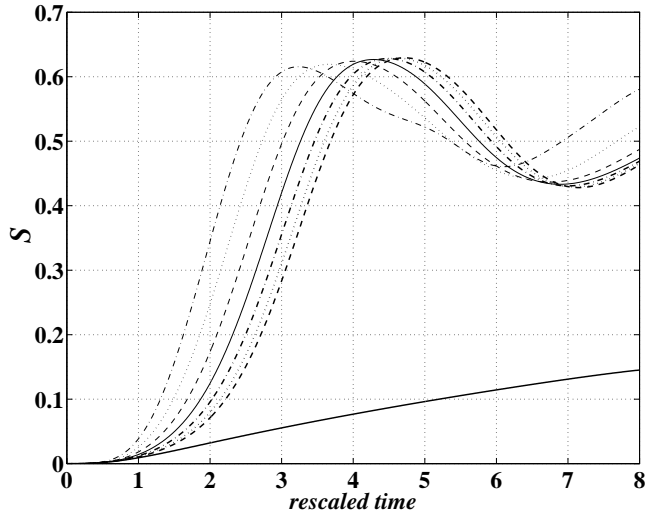,width=0.8\columnwidth,angle=90}
\end{center}
\caption{Growth of the von Neumann entropy $S$ of the quantum reduced
single-particle density operator in the presence of thermal noise (
$\Gamma=10^{-2}\Omega$), at $\kappa=2\Omega$, for N=10 ($-\cdot-$),
20 ($\cdot\cdot\cdot$), 40 ($---$), 80 (------), 160 ({\bf - . - }), 
320 ({\bf - - - -}, and 640 ({\bf . . .}) particles. Bold solid curve corresponds
to the mean field entropy. Initial conditions are the same as in Fig. 3. }
\label{f6}
\end{figure}
In Fig. 7, we summarize the results of numerous dynamical calculations
conducted for various values of the particle number $N$ and of the thermal
noise $\Gamma $, by plotting the time at which the entropy reaches a given
value. The curves are obtained using the modified BBR equations (\ref{bbrdec}%
) whereas the circles and squares depict exact quantum results (limited by
computation power to $N\sim 10^{3}$ particles) for two limiting values of $%
\Gamma $. The BBR equations provide accurate predictions of the initial
decoherence rate and the quantum break time even within this limited range
of $N$ (and the agreement between the exact quantum results and the BBR
predictions would become still better for higher numbers of particles). Once
more, we observe the logarithmic growth of the quantum break time with $N$
in the zero temperature ($\Gamma =0$) limit. However, when the temperature
is finite, there is a saturation of the quantum break time to values which
are well below the mean-field thermal dephasing times, in agreement with
Fig. 6. 

Instead of observing the quantum break time as a function of the number of
particles for a given degree of thermal noise, we can monitor the thermal
decoherence time as a function of temperature, for any given number of
particles. Viewing Fig. 7 this way, it is evident that in the mean-field
limit ($1/\sqrt{N}\rightarrow 0$) the purely thermal dephasing time also
grows only logarithmically with the temperature, Comparing this result to
the  $\log (N)$ growth of the quantum break time in the zero-temperature
limit, we can see that thermal noise and quantum noise have essentially
similar effects on the system. \ And Figs. 6 and 7 together are in complete
agreement with the prediction that the entropy of a dynamically unstable
quantum system coupled to a reservoir\cite{PazZurek}, or of a stable system
coupled to a dynamically unstable reservoir, will grow linearly with time,
at a rate independent of the system-reservoir coupling, after an onset time
proportional to the logarithm of the coupling \cite{mohanty,pattanayak}.
Thus, one can really consider the Bogoliubov fluctuations as a reservoir 
\cite{habib}, coupled to the mean field with a strength proportional to $1/N$%
. \ The $N\leftrightarrow 1/T$ analogy is even further extended by the
saturation for any finite $N$, of the thermal dephasing time at low $T$, in
the same way that the quantum break time for a finite $T$ saturates at high $%
N$. \ Due to this quantum saturation, quantum corrections can be
experimentally distinguished from ordinary thermal effects which do not
saturate the dephasing rate at low temperature. \  
\begin{figure}
\begin{center}
\epsfig{file=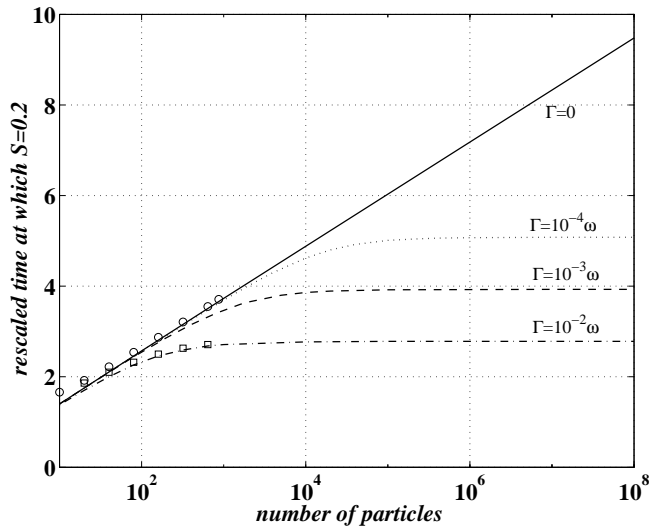,width=\columnwidth}
\end{center}
\caption{ Time at which $S$ reaches 0.2 as a function of the particle number 
$N$, according to the BBR equations (\ref{bgb}), modified to include thermal
phase-diffusion. Four different values of $\Gamma $ are shown: $\Gamma =0$
(-----), $\Gamma =10^{-4}\Omega $ ($\cdot \cdot \cdot $), 
$\Gamma=10^{-3}\Omega $ ($---$), and $\Gamma =10^{-2}\Omega $ ($-\cdot -$). Exact
quantum results are presented for $\Gamma =0$ (circles) and $\Gamma
=10^{-2}\Omega $ (squares). Initial conditions, $\protect\kappa $ and 
$\Omega $ are the same as in Fig. 6.}
\label{f7}
\end{figure}
\section{Scattering-length measurements}

After indicating how condensate decoherence at dynamical instabilities can
connect principles established in different areas of physics, we briefly
note that it can also have practical applications. \ Rapid decoherence in
the vicinity of the unstable $\pi $-state of the two-mode condensate may
serve for the direct measurement of scattering lengths. As demonstrated in
Fig. 8, the mean-field trajectory of a condensate which is prepared
initially in one of the modes, would only pass through the rapidly dephasing
unstable point when $\kappa=2\Omega $. Thus, the self-interactions energy $%
\kappa $ can be determined by measuring the entropy at a fixed time as a
function of the coupling frequency $\Omega $, resulting in a sharp line
about $\Omega =\kappa/2$, as depicted in Fig. 9.

Experimentally, the single-particle entropy is measurable, in the internal
state realization of our model, by applying a fast Rabi pulse and measuring
the amplitude of the ensuing Rabi oscillations, which is proportional to the
Bloch vector length $|{\bf {\vec s}}|$. (Successive measurements with Rabi
rotations about different axes, i.e. by two resonant pulses differing by a
phase of $\pi/2$, will control for the dependence on the angle of ${\bf {%
\vec s}}$). In a double well realization, one could determine the
single-particle entropy by lowering the potential barrier, at a moment when
the populations on each side were predicted to be equal, to let the two
parts of the condensate interfere. The fringe visibility would then be
proportional to $|{\bf {\vec s}}|$ \cite{ketterle}. 
\begin{figure}
\begin{center}
\epsfig{file=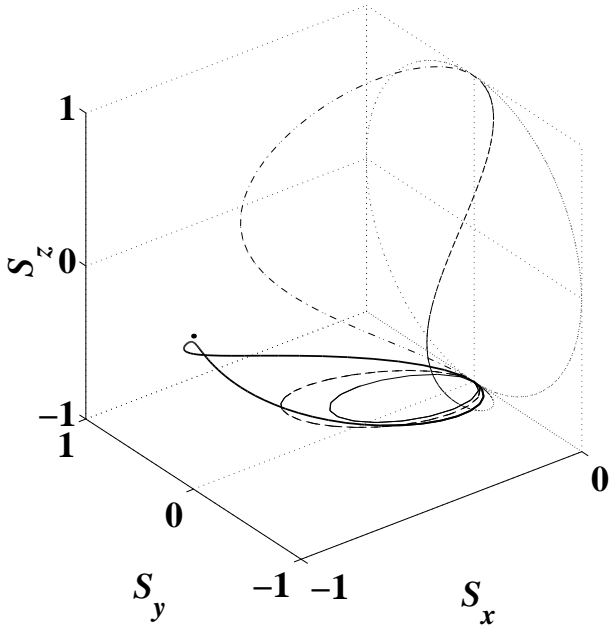,width=\columnwidth,angle=90}
\end{center}
\caption{Mean-field trajectories starting at ${\bf {\vec{s}}}=(0,0,-1)$ for 
$\kappa =0$ ($\dots $), $\protect\kappa =\Omega $ ($-\cdot -$), 
$\kappa =2\Omega $ ({\bf -----}), $\kappa =3\Omega $ ($---$),
and $\protect\kappa =4\Omega $ (-----). The dot at ${\bf {\vec{s}}}=(-1,0,0)$
marks the dynamical instability.}
\label{f8}
\end{figure}
\section{Conclusions}

To conclude, we have shown that significant quantum corrections to the
Gross-Pitaevskii MFT, in the vicinity of its dynamical instabilities, can be
measured in a two-mode BEC under currently achievable experimental
conditions. We have derived a simple theory that accurately predicts the
leading quantum corrections and the quantum break time. \ By applying to
condensate physics some insights from studies of decoherence, we have found
evidence that MFT dynamical instabilities cause linear growth of the
single-particle entropy at a rate independent of $N.$ \ And from condensate
physics we have learned something about decoherence: we have identified a
form of decoherence which degrades quantum-classical correspondence, instead
of improving it. \ 
\begin{figure}
\begin{center}
\epsfig{file=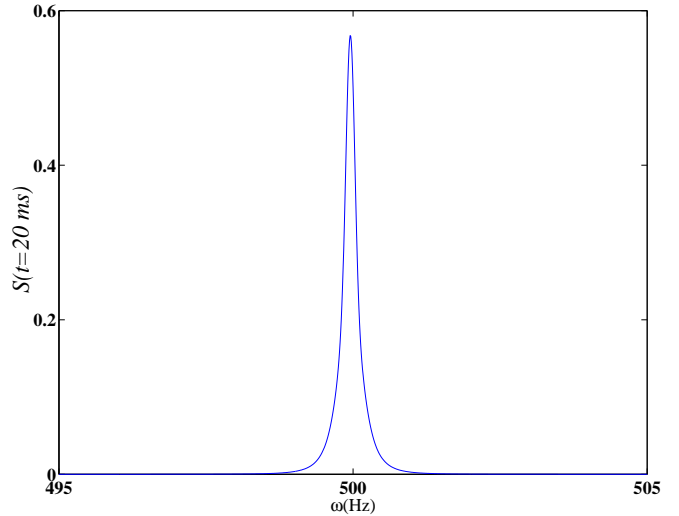,width=0.8\columnwidth,angle=90}
\end{center}
\caption{von Neumann entropy $S$ after 20 ms of propagation according to the
BBR equations (\ref{bbrdec}) with $\protect\kappa=1$ KHz and $\Gamma=10^{-4}$
Hz, starting with the entire condensate of $N=10^5$ particles in one mode,
as a function of the coupling frequency $\Omega$.}
\label{f9}
\end{figure}
Our picture of quantum backreaction in BECs as decoherence suggests new
lines of investigation for both experiment and theory: measurements of
single-particle entropy in condensates, descriptions of condensates with
mixed single particle states (instead of the usual macroscopic wave
functions), and general questions of decoherence under nonlinear evolution.
Exploring these possibilities, beyond the two-mode model considered here,
provides many goals for further research.

\section*{Acknowledgments}
This work was supported by the National Science Foundation through a 
grant for the Institute for Theoretical Atomic and Molecular Physics 
at Harvard University and Smithsonian Astrophysical Observatory.

\newpage

\begin{references}
\bibitem{CastinDum}  Y.~Castin and R.~Dum, Phys. Rev. Lett. {\bf 79}, 3553
(1997).  

\bibitem{vardi}  A. Vardi and J. R. Anglin, Phys.~Rev.~Lett.~, {\bf 86}, 568 
(2001).

\bibitem{javanainen}  J. Javanainen, Phys.~Rev.~Lett.~{\bf 57},\ 3164\
(1986); J. Javanainen and S. M. Yoo, Phys.~Rev.~Lett.~{\bf 76},\ 161\ (1996).

\bibitem{walls}  M. W. Jack, M. J. Collett, and D. F. Walls, 
Phys.~Rev.~A~{\bf 54},\
R4625\ (1996); G. J. Milburn, J. Corney, E. M. Wright, D. F. Walls,
Phys.~Rev.~A~{\bf 55},\ 4318\ (1997); A. S. Parkins and D. F. Walls
Phys.~Rep. ~{\bf 303},\ 1\ (1998).

\bibitem{Janne}  J. Ruostekoski and D. F. Walls, Phys.~Rev.~A~{\bf 58},\
R50\ (1998).

\bibitem{smerzi}  A. Smerzi, S. Fantoni, S. Giovanazzi, and S. R. Shenoy,
Phys.~Rev.~Lett.~{\bf 79},\ 4950\ (1997); S. Raghavan, A. Smerzi, S.
Fantoni, and S. R. Shenoy, Phys.~Rev.~A~{\bf 59},\ 620\ (1999); I. Marino,
S. Raghavan, S. Fantoni, S. R. Shenoy, and A. Smerzi, Phys.~Rev.~A~{\bf 60}%
,\ 487\ (1999).

\bibitem{leggett}  I. Zapata, F. Sols, and A. J. Leggett, Phys.~Rev.~A~{\bf 57}%
,\ R28\ (1998).

\bibitem{lewenstein}  P. Villain and M. Lewenstein, Phys.~Rev.~A~{\bf 59},\
2250\ (1999).

\bibitem{ketterle}  M. R. Andrews, C. G. Townsend, H. J. Miesner, D. S.
Durfee, D. M. Kurn, and W. Ketterle, Science~{\bf 275},\ 637\ (1997).

\bibitem{cornell1}  M. R. Matthews, B. P. Anderson, P. C. Haljan, D. S.
Hall, M. J. Holland, J. E. Williams, C. E. Wieman, and E. A. Cornell,
Phys.~Rev.~Lett.~{\bf 83},\ 3358\ (1999).

\bibitem{cornell2}  C. J. Myatt, E. A. Burt, R. W. Ghrist, E. A. Cornell,
and C. E. Wieman, Phys.~Rev.~Lett.~{\bf 78},\ 586\ (1997).

\bibitem{cornell3}  M. R. Matthews, D. S. Hall, D. S. Jin, J. R. Ensher, C.
E. Wieman, E. A. Cornell, F. Dalfovo, C. Minniti, and S. Stringari,
Phys.~Rev.~Lett.~{\bf 81},\ 243\ (1998); J. Williams, R. Walser, J. Cooper,
E. Cornell, and M. Holland, Phys.~Rev.~A~{\bf 59},\ R31\ (1999).

\bibitem{stenholm}  Patrik \"Ohberg and Stig Stenholm, Phys.~Rev.~A~{\bf 59}%
,\ 3890\ (1999).

\bibitem{griffin}  A. Griffin, Phys.~Rev.~B~{\bf 53},\ 9341\ (1996).

\bibitem{giulini}  See e.g. D. Giulini, E. Joos, C. Kiefer, J. Kupsch, I.-O.
Stamatescu, and D. Zeh, {\it Decoherence and the Appearance of a Classical
World in Quantum Theory} (Springer, Berlin, 1996).

\bibitem{Anglin}  J. R.~Anglin, Phys. Rev. Lett. {\bf 79}, 6 (1997).

\bibitem{PazZurek}  J.-P.~Paz and W.H.~Zurek, Phys. Rev. Lett. {\bf 72},
2508 (1994).

\bibitem{mohanty}  P. Mohanty, E. M. Q. Jariwala, and R. A. Webb, 
Phys.~Rev.~Lett.~{\bf 78}%
,\ 3366\ (1997).

\bibitem{pattanayak}  A. K. Pattanayak and P. Brumer, Phys.~Rev.~Lett.~{\bf %
79},\ 4131\ (1997).

\bibitem{habib}  S. Habib, Y. Kluger, E. Mottola, and J.-P. Paz, Phys. Rev.
Lett. {\bf 76}, 4660 (1996).
\end{references}
\end{document}